\documentclass[lettersize,journal]{IEEEtran}

\usepackage{inputenc}
\usepackage[style=ieee, backend=biber, isbn=false]{biblatex}

\usepackage{amsmath,amsfonts}
\usepackage{algorithmic}
\usepackage{algorithm}
\usepackage{array}
\usepackage[caption=false,font=normalsize,labelfont=sf,textfont=sf]{subfig}
\usepackage{textcomp}
\usepackage{stfloats}
\usepackage{url}
\usepackage{verbatim}
\usepackage{graphicx}
\usepackage{todonotes}
\usepackage{comment}

\usepackage{amssymb}
\usepackage{latexsym}
\usepackage{sansmath}

\usepackage{url}
\usepackage{xcolor}
\definecolor{newcolor}{rgb}{.8,.349,.1}

\usepackage{comment}

\usepackage{import}

\usepackage[utf8]{luainputenc}

\usepackage[T1]{fontenc}

\usepackage{graphicx}

\usepackage[english]{babel}
\addto\captionsenglish{}
\addto\captionsenglish{}
\usepackage{csquotes}

\usepackage{newtxtext}
\usepackage{amsthm}
\usepackage[slantedGreek]{newtxmath}
\usepackage[OMLmathsfit]{isomath}
\DeclareMathAlphabet{\mathbfsf}{\encodingdefault}{\sfdefault}{bx}{n}
\usepackage{bm}
\usepackage{envmath}
\usepackage{mathtools}
\usepackage{commath}
\usepackage{siunitx}

\usepackage{booktabs}
\usepackage{footmisc}  

\usepackage{url}

\theoremstyle{definition}

\theoremstyle{plain}

\theoremstyle{remark}

\usepackage{lineno}
\modulolinenumbers[5]
\usepackage{umoline}

\usepackage{pgfplots}
\usepackage{pgfplotstable}
\pgfplotsset{compat=newest}
\pgfplotsset{plot coordinates/math parser=false}
\newlength\figureheight
\newlength\figurewidth
\pgfplotsset{every axis plot/.append style={line width=1.5pt},
    legend style={font=\footnotesize, 
        text height=1.0ex,
        draw=black,
        fill=white,
        legend cell align=left}}

\usepackage{hyperref} 
\usepackage[english]{cleveref}

\Crefname{defn}{definition}{definitions}
\Crefname{defn}{Definition}{Definitions}

\Crefname{asm}{assumption}{assumptions}
\Crefname{asm}{Assumption}{Assumptions}

\crefname{lem}{lemma}{lemmas} 
\Crefname{lem}{Lemma}{Lemmas}

\crefname{prop}{proposition}{propositions} 
\Crefname{prop}{Proposition}{Propositions}

\crefname{thm}{theorem}{theorms} 
\Crefname{thm}{Theorem}{Theorms}

\crefname{cor}{corollary}{corollaries}
\Crefname{cor}{Corollary}{Corollaries}
\newcounter{subequation}
\newlength\mtabskip\mtabskip=-1.25cm

\def\mtabLong{long}

\newcommand{\mr}{\mathrm}

\newcommand{\veg}[1]{\bm{#1}}     
\newcommand{\mat}[1]{\mathsfbfit{#1}} 
\renewcommand{\vec}[1]{\mathsfbfit{#1}} 
\newcommand{\op}[1]{\mathcal{#1}} 
\newcommand{\vecop}[1]{\bm{\mathcal{#1}}} 

\DeclareMathOperator{\ee}{e}

\newcommand{\T}{\mr{T}}
\newcommand{\Ht}{\dagger}

\newcommand\restr[2]{{
        \left.\kern-\nulldelimiterspace 
        #1 
        \vphantom{|} 
        \right|_{#2} 
}}

\newcommand\rst[3]{{
        \left.\kern-\nulldelimiterspace 
        #1 
        \vphantom{|} 
        \right|_{#2}^{#3} 
}}

\usepackage{acro}

\DeclareAcronym{DG}
{
    short = DG ,
    long = discontinuous Galerkin
}

\DeclareAcronym{ACA}
{
    short = ACA ,
    long = adaptive cross approximation
}

\DeclareAcronym{EFIE}
{
    short =  EFIE ,
    long = electric field integral equation
}

\DeclareAcronym{MFIE}
{
    short =  MFIE ,
    long = magnetic field integral equation
}

\DeclareAcronym{CFIE}
{
    short =  CFIE ,
    long = combined field integral equation
}

\DeclareAcronym{MUIE}
{
    short =  MUIE ,
    long = Müller integral equation
}

\DeclareAcronym{PMCHWT}
{
    short =  PMCHWT ,
    long = Poggio-Miller-Chang-Harrington-Wu-Tsai integral equation
}

\DeclareAcronym{SPD}
{
    short =  SPD ,
    long = {symmetric, positive definite}
}

\DeclareAcronym{SPSD}
{
    short =  SPD ,
    long = {symmetric, positive semi-definite}
}

\DeclareAcronym{PEC}
{
    short =  PEC ,
    long = perfectly electrically conducting
}

\DeclareAcronym{RWG}
{
    short = RWG ,
    long = Rao-Wilton-Glisson
} 

\DeclareAcronym{BC}
{
    short = BC ,
    long = Buffa-Christiansen
}

\DeclareAcronym{SVD}
{
    short = SVD ,
    long = singular value decomposition
}

\DeclareAcronym{CG}
{
    short = CG ,
    long = conjugate gradient
} 

\DeclareAcronym{PCG}
{
    short = PCG ,
    long = preconditioned conjugate gradient
} 

\DeclareAcronym{CGS}
{
    short = CGS ,
    long = conjugate gradient squared
}

\DeclareAcronym{CMP}
{
    short = CMP ,
    long = Calderón multiplicative preconditioner
} 

\DeclareAcronym{RFCMP}
{
    short = RF-CMP ,
    long = refinement-free Calderón multiplicative preconditioner
} 

\DeclareAcronym{HPD}
{
    short = HPD ,
    long = {Hermitian, positive definite}
} 

\DeclareAcronym{RHS}
{
    short = RHS ,
    long = {right-hand side}
}

\DeclareAcronym{PW}
{
    short = PW ,
    long = {plane wave}
} 

\DeclareAcronym{HD}
{
    short = HD ,
    long = {Hertzian dipole}
} 

\DeclareAcronym{FF}
{
    short = FF ,
    long = {far-field}
} 

\DeclareAcronym{NF}
{
    short = NF ,
    long = {near-field}
}  

\newcolumntype {n}{c}
\newcolumntype {N}{>{\small}c}
\newcolumntype {L}{>{\small}l}
\newcolumntype {F}{>{\footnotesize}c}
\newcolumntype {v}[1]{>{\raggedright \hspace {0pt}} p {#1}}
\newcolumntype {V}[1]{>{\small \raggedright \hspace {0pt}} p {#1}}
\newcolumntype{d}[1]{>{\DC@{.}{.}{#1}}c<{\DC@end}}

\newcolumntype{R}[1]{%
    >{\begin{turn}{90}\begin{minipage}{#1}\small\raggedright\hspace{0pt}}l%
            <{\end{minipage}\end{turn}}%
}

\NewDocumentCommand{\TA}{o}{
    \IfNoValueTF {#1} {%
        \vecop T_{\kern-2pt\mr{A}}
    }
    {
        \vecop T_{\kern-2pt\mr{A},#1}
    }
}

\NewDocumentCommand{\TPhi}{o}{
    \IfNoValueTF {#1} {%
        \vecop T_{\kern-2pt\Phiup}
    }
    {
        \vecop T_{\kern-2pt\Phiup,#1}
    }
}

\NewDocumentCommand{\matTA}{o}{
    \IfNoValueTF {#1} {%
        \mat T_\mr{A}   
        }
    {
        \mat T_{\mr{A},#1}
    }
}

\NewDocumentCommand{\matTPhi}{o}{
    \IfNoValueTF {#1} {%
        \mat T_\Phiup   
        }
    {
        \mat T_{\Phiup,#1}
    }
}

\NewDocumentCommand{\MSL}{o}{
    \IfNoValueTF {#1} {%
        \veg \Psi_\mr{SL}
        }
    {
        \veg \Psi_{\mr{SL},#1}
    }
}

\NewDocumentCommand{\MDL}{o}{
    \IfNoValueTF {#1} {%
        \veg \Psi_\mr{DL}
        }
    {
        \veg \Psi_{\mr{DL},#1}
    }
}

\NewDocumentCommand{\PA}{o}{
    \IfNoValueTF {#1} {%
        \veg \Psi_\mr{A}
        }
    {
        \veg \Psi_{\mr{A},#1}
    }
}

\NewDocumentCommand{\PPhi}{o}{
    \IfNoValueTF {#1} {%
        \veg \Psi_{\Phiup}
        }
    {
        \veg \Psi_{\Phiup,#1}
    }
}

\begin{document}

\newcommand{\jj}{\mathrm{j}}

\title{\fontsize{24}{28}\selectfont Spectral Analysis of Discretized Boundary Integral Operators in 3D: a High-Frequency Perspective}


\author{
	\IEEEauthorblockN{Viviana Giunzioni\IEEEauthorrefmark{1}, Adrien Merlini\IEEEauthorrefmark{2}, Francesco P. Andriulli\IEEEauthorrefmark{1}}
	
\IEEEauthorblockA{\IEEEauthorrefmark{1} Department of Electronics and Telecommunications, Politecnico di Torino, 10129 Turin, Italy }

\IEEEauthorblockA{\IEEEauthorrefmark{2} Microwave Department, IMT Atlantique, 29238 Brest, France}
	}



\pagenumbering{gobble}

\maketitle

\begin{abstract}
When modeling propagation and scattering phenomena using integral equations discretized by the boundary element method, it is common practice to approximate the boundary of the scatterer with a mesh comprising elements of size approximately equal to a fraction of the wavelength $\lambda$ of the incident wave, e.g., $\lambda/10$. In this work, by analyzing the spectra of the operator matrices, we show a discrepancy with respect to the continuous operators which grows with the simulation frequency, challenging the common belief that the aforementioned widely used discretization approach is sufficient to maintain the accuracy of the solution constant when increasing the frequency.   
\end{abstract}

\begin{IEEEkeywords}
Integral equations, spectral analysis, high-frequency, discretization error.
\end{IEEEkeywords}

\section{Introduction}
\IEEEPARstart{B}{oundary} element methods (BEMs) for discretizing and solving integral equations are fundamental tools in modeling wave propagation and scattering phenomena.
When increasing the simulation frequency $k$, one common approach consists in refining the discretization of the problem, increasing the number of degrees of freedom proportionally to $k^{d-1}$, where $d$ is the dimension of the problem. This is equivalent, for example, to employ a mesh approximating the geometry characterized by elements of size $h$ approximately equal to a fraction of the wavelenght. The regime characterized by a constant value of $hk$ when $k$ goes to infinity is referred to as high-frequency regime. Although this kind of refinement is often the preferred choice, its efficacy in keeping the solution error constant in frequency is still topic of discussion \cite{galkowski2019wavenumberexplicit,giunzioni2024highfrequency}.

In this contribution, we propose a new approach to explore this fundamental question based on the spectral analysis of boundary integral operators in the high-frequency regime.
To this end, we develop a novel semi-analytic approach to determine the eigenvalues of the matrices discretizing the operators over the sphere, which can then be used to show how the discretization process causes the spectra of the matrices and of the continuous operators from which they derive to differ.

Once the eigenvalues of the BEM matrices are available, they can be compared to the ones of the continuous operators the matrices discretize, which results in the definition of a spectral relative difference, whose behavior in the high-frequency limit is studied in this contribution.
Our analysis leads to the conclusion that the BEM discretization of one of the operators considered is affected by a spectral error increasing in the high-frequency regime, challenging the common belief that the accuracy is constant in said regime.

The paper is organized as follows: after providing the required notation in Section~\ref{sec:background}, we detail in Section~\ref{sec:eigenvalues} our new approach for the evaluation of the eigenvalues of the operator matrices. Subsequently, we propose a high-frequency analysis of the spectral error between matrices and operators in Section~\ref{sec:analysis}. Some numerical results proposed in Section~\ref{sec:results} will validate the proposed approach, enforcing the conclusions of this work, drawn in Section~\ref{sec:conclusion}.

\section{Background and Formalism}
\label{sec:background}

Given the three-dimensional, free-space Green's function of wavenumber $k$
\begin{equation}
    G^k( \veg \rho,  \veg\rho') \coloneqq \frac{e^{-\jj k | \veg\rho-  \veg\rho'|}}{4 \pi |\veg \rho-  \veg\rho'|}\,,
\end{equation}
and a closed and orientable surface $\Gamma$ characterized by the outward normal $\veg n$,
we define the single-layer and hypersingular operators as
\begin{align}
    \op S^k f(\veg \rho) &\coloneqq k \int_\Gamma G^k (\veg \rho,\veg \rho') f(\veg \rho') d \veg \rho'\,,\\
    \op N^k f(\veg \rho) &\coloneqq - \frac{1}{k}\frac{\partial}{\partial \veg n} \int_\Gamma \frac{\partial}{\partial  \veg n'} G^k (\veg \rho,\veg \rho') f(\veg \rho') d \veg \rho'\,.
\end{align}

In this work, we assume that $\Gamma$ is the sphere of radius $a$ centered in the origin. In this case, the eigenvalue decompositions of the above operators are known \cite{hsiao1994error},
\begin{equation}
    \op S^k Y_l^p = \lambda_l^{\op S^k} Y_l^p\,\,,\quad  \op N^k Y_l^p = \lambda_l^{\op N^k} Y_l^p\,,
\end{equation}
where $Y_l^p = Y_l^p(\theta,\phi)$ denotes the spherical harmonic function of the polar and azimuthal coordinates defined as in \cite[Section 8.1]{abramowitz1964handbook},
and the associated eigenvalues $\lambda_l^{\op S^k}$ and $\lambda_l^{\op N^k}$, of multiplicity $2l+1$, have expressions
\begin{align}
    \lambda_l^{\op S^k} &= -\jj (ka)^2 j_l(ka) h_l^{(2)}(ka)\,,\\
    \lambda_l^{\op N^k} &= \jj (ka)^2 j_l^\prime(ka) h_l^{(2)\prime}(ka)\,,
\end{align}
where $j_l$ and $h_l^{(2)}$ are the spherical Bessel function of the first kind and the spherical Hankel function of the second kind respectively, of order $l$ \cite{abramowitz1964handbook}.

Boundary element approaches to discretize the integral equations involving $\op S^k$ and $\op N^k$ commonly rely on the definition of a set of testing $t$ and source $f$ functions defined over a discretization or parametrization of $\Gamma$ and on the definition of the matrices $\mat S^k$ and $\mat N^k$ with elements proportional to
\begin{equation}
    (\mat S^k)_{uv} \propto \left( t_u , \op S^k f_v \right)_{L^2(\Gamma)}\,\,,\quad
    (\mat N^k)_{uv} \propto \left( t_u , \op N^k f_v \right)_{L^2(\Gamma)}\,.\nonumber
\end{equation}
In the next section, we aim at evaluating the eigenvalues of these matrices in case of source and test functions defined on a partition of the sphere.

\section{A New Semi-Analytic Approach to Evaluate the Eigenvalues of Matrices $\mat S^k$ and $\mat N^k$}
\label{sec:eigenvalues}
We define a partition of $\Gamma$, i.e., a set of $V^2$ surfaces $\Gamma_i$ such that $\cup_{i=1}^{V^2} \Gamma_i = \Gamma$. Differently from standard triangulations or higher-order discretizations of surfaces commonly employed in BEM approaches, this choice leads to a nonexistent geometrical discretization error. After introducing the standard spherical coordinates $\phi \in [0,2\pi)$ and $\theta\in [ 0,\pi ]$, representing the azimuthal and polar angle, we propose a uniform discretization along $\phi$ and $z\coloneqq\cos(\theta)$, as $\phi_\mu =  \mu h_\phi$ for $\mu = 0,\dots,V-1$ and $z_\nu = 1-\nu h_z$ for $\nu = 0,\dots,V$, where $h_\phi = 2\pi/V$ and $h_z = 2a/V$; $V$ is assumed to be odd. The $V^2$ portions of $\Gamma$ characterized by $\phi_\mu\le\phi\le \phi_{\mu+1}$ and $\theta_{\nu}\le\theta\le \theta_{\nu+1}$ have an equal area $A\coloneqq 4\pi a^2/V^2$ and will be employed as a partition of $\Gamma$. 

To discretize $\op S^k$ and $\op N^k$ using the BEM, we use test and source basis functions defined over the spherical domain $\Gamma$, which are separable functions of $\phi$ and $z$. In particular, after defining two sets of $M=V$ functions along $\phi$, 
\begin{equation}
    \{t_m^\phi(\phi)\}_{m=0}^{M-1}\quad \text{and}\quad \{f_m^\phi(\phi)\}_{m=0}^{M-1}\,,
\end{equation}
and two sets of $N=V$ functions along $z$,
\begin{equation}
    \{t_n^z(z)\}_{n=0}^{N-1}\quad \text{and}\quad\{f_n^z(z)\}_{n=0}^{N-1}\,,
\end{equation}
we can define two sets of $(NM)=V^2$ test and source functions
\begin{align}
    t_{nM+m}(z,\phi) &= t^z_n(z) t^\phi_m(\phi)\,,\label{eqn:indexing1}\\
    f_{nM+m}(z,\phi) &= f^z_n(z) f^\phi_m(\phi)\,.\label{eqn:indexing2}
\end{align}
Additionally, we further assume shift-invariance of $t_m^\phi$ and $f_m^\phi$,
\begin{equation}
    t_m^\phi(\phi) = t_0^\phi(\phi-\phi_m)\,\,,\quad f_m^\phi(\phi) = f_0^\phi(\phi-\phi_m)\,.
\end{equation}

The matrix $\mat S^k$ resulting from the BEM discretization of the single-layer operator has elements defined as
\begin{equation}
    \left(\mat S^k\right)_{uv} = \frac{k}{A} \int_\Gamma t_u(\veg r) \int_\Gamma f_v(\veg r') G^k(\veg r,\veg r')\, \mathrm{d} \veg r'\, \mathrm{d} \veg r\,.
    \label{eqn:Selementdefinition}
\end{equation}
Using the addition theorem for spherical harmonics and spherical Bessel functions, the Green's function evaluated at points $\veg r = (\theta,\phi)$ and $\veg r' = (\theta',\phi')$ can be expanded in as \cite[Eq. 3.7.4]{chew1995waves}
\begin{equation}
    G^k(\veg r,\veg r') = -\frac{1}{k a^2}\sum_{l=0}^\infty \sum_{p=-l}^l Y_l^p(\theta,\phi) Y_l^{p*}(\theta',\phi') \lambda_l^{\op S^k}\,.
    \label{eqn:additiontheorem}
\end{equation}
By substituting \eqref{eqn:additiontheorem} in \eqref{eqn:Selementdefinition}, given the separability of the basis functions, after a change of variables from Cartesian to spherical coordinates, the element of $\mat S^k$ can be written as
\begin{gather}
    \left(\mat S^k\right)_{nM+m,n^\prime M+m^\prime} = \nonumber\\
    -\sum_{l=0}^\infty \sum_{p=-l}^l \lambda_l^{\op S^k} 
    \left(\int_{-a}^a \bar{P}_l^p(z) t^z_n({z})\, d z\right)
    \left(\int_{-a}^a \bar{P}_l^p(z') f^z_{n^\prime}(z')\, d z'\right)\nonumber\\
     \left(\int_{0}^{2\pi} \ee^{\mathrm{j}p\phi} t^\phi_m(\phi)\, d \phi\right)
     \left(\int_{0}^{2\pi} \ee^{-\mathrm{j}p\phi'} f^\phi_{m^\prime}(\phi')\, d \phi'\right)\frac{V^2}{4\pi a^4}\,,
\end{gather}
where we denote by $\bar{P}_l^p(z)$ the normalization of the Legendre function as $\bar{P}_l^p(z) = \sqrt{(2l+1)(l-|p|)!/(l+|p|)!}\,{P}_l^p(z)$.

The indexing of basis functions introduced in \eqref{eqn:indexing1}, \eqref{eqn:indexing2} makes the matrix $\mat S^k$ block-circulant, since it is composed of $N^2$ circulant blocks of dimension $M$. Indeed, each of these blocks corresponds to the interaction between test functions defined over a spherical segment
and source functions defined over another spherical segment.
We now define the matrix $\mat D_M\in\mathbb{C}^{M\times M}$ whose $(p+\lfloor M/2\rfloor)-$th column has its $m-$th element equal to $\ee^{-\mathrm{j} p \phi_m}/\sqrt{M}$; this matrix is the matrix of eigenvectors of each of the circulant blocks of $\mat S^k$ \cite{briggs1995dft,warnick2008numerical}. We also define $\mat D\in\mathbb{C}^{MN\times MN}$ as a block diagonal matrix with diagonal blocks equal to $\mat D_M$. The result of the matrix product $\mat D^{\Ht} \mat S^k \mat D$ is a block matrix containing $N^2$ diagonal blocks of dimension $M$. The diagonal of each of these blocks contains the eigenvalues of the corresponding block of $\mat S^k$, of spectral index $p$ ranging from $-\lfloor M/2 \rfloor$ to $\lfloor M/2 \rfloor$. Then, by means of the permutation matrix $\mat P$, we apply a permutation aimed at grouping together all the eigenvalues of the different blocks corresponding to the same eigenvector, or, equivalently, characterized by the same spectral index $p$. The resulting matrix $\bar{\mat S}^k$ is a block diagonal matrix, in which the block of index $(p+\lfloor M/2 \rfloor)$ contains in index $(i,j)$ the eigenvalue of spectral index $p$ of the block of index $(i,j)$ of matrix $\mat S^k$ \cite{franzo2024fast} 
\begin{equation}
    \mat P^{\T} \mat D^{\Ht} \mat S^k \mat D \mat P = \bar{\mat S}^k\,.
\end{equation}
Given the diagonalization of the original and of the manipulated matrices,
\begin{align}
    \mat S^k \vec v &= \lambda \vec v\\
    \mat P^{\T} \mat D^{\Ht} \mat S^k \mat D \mat P \bar{\vec v} &= \bar{\lambda} \bar{\vec v}\,,
\end{align}
we have that
\begin{equation}
    \bar{\vec v} = \mat P^{\T}\mat D^{\Ht} \vec v 
    \quad \text{and} \quad \bar{\lambda} =
    \lambda\,,
\end{equation}
that is, matrices $\mat S^k$ and $\bar{\mat S}^k$ share the same eigenvalues. As a consequence, it is possible to retrieve the eigenvalues of $\mat S^k$ as the union of the eigenvalues of the diagonal blocks of $\bar{\mat S}^k$.
To facilitate further developments we denote the diagonal block of $\bar{\mat S}^k$ at index $(p+\lfloor M/2 \rfloor)$ as $\bar{\mat S}^k_p$, and the Fourier series coefficients of $t^\phi_0(\phi)$ and $f^\phi_0(\phi)$ as $T_p$, $F_p$ \cite{warnick2008numerical},
\begin{equation}
    T_p \coloneqq \frac{1}{h_\phi}\int t^\phi_0(\phi) \ee^{\jj p \phi}\,d\phi\,,\,\,
    F_p \coloneqq \frac{1}{h_\phi}\int f^\phi_0(\phi) \ee^{\jj p \phi}\,d\phi\,.
\end{equation}
The matrix $\bar{\mat S}^k_p$ can be expressed as
\begin{align}
    \bar{\mat S}^k_p &= -\frac{V}{2 a^2} \sum_{l=0}^\infty  \lambda_l^{\op S^k} \sum_{\substack{s=-\infty \\ |p+sM|\le l}}^\infty  \mat P_l^{(p+sM)} T_{-(p+sM)}F_{(p+sM)}\nonumber\\
    &= -\frac{V}{2 a^2} \sum_{s=-\infty}^\infty  \sum_{l=|p+sM|}^\infty \lambda^{\op S^k}_l \mat P_l^{(p+sM)} T_{-(p+sM)}F_{(p+sM)}\,,\label{eqn:doublesum}
\end{align}
where matrix $\mat P_l^p$ is a rank-one matrix, $\mat P_l^p = \vec t_l^p \vec f_l^{p \T}$, with
\begin{equation}
    \vec t_l^p = \begin{pmatrix}
        \int_{-a}^a \bar{P}_l^{p}(z)t^{z}_0(z) dz\\
        \vdots\\
        \int_{-a}^a \bar{P}_l^{p}(z)t^{z}_{N-1}(z) dz
    \end{pmatrix} \,,\,\,
     \vec f_l^p = \begin{pmatrix}
        \int_{-a}^a \bar{P}_l^{p}(z)f^{z}_0(z) dz\\
        \vdots\\
        \int_{-a}^a \bar{P}_l^{p}(z)f^{z}_{N-1}(z) dz
    \end{pmatrix}\,.
    \nonumber
\end{equation}

Using these results, the eigenvalues of $\mat S^k$ can be compared with the eigenvalues of the operator $\op S^k$. We focus in particular on the hyperbolic and the transition regions of the spectrum, corresponding respectively to spectral indices $l \ll ka$ and $l\simeq ka$ \cite{giunzioni2024highfrequency}. Given the nature of the summation with respect to the index $l$ in \cref{eqn:doublesum}, the eigenvalues of the block $\bar{\mat S}^k_{p=0}$ should be compared with $\lambda_l^{\op S^k}$ for $l=0,\dots,V-1$, where, given the multiplicity of the eigenvalues of the operators, $V-1\eqcolon l_{\text{max}}$ is the maximum spectral index of the visible range. Indeed, ${\sum_{l=0}^{l_{\text{max}}}(2l+1)}=(l_{\text{max}}+1)^2=V^2$. Similarly, $(V-1)$ eigenvalues of $\bar{\mat S}^k_{p=\pm 1}$ should be compared with $\lambda_l^{\op S^k}$ for $l=1,...,V-1$, while one remaining eigenvalue does not have clear correspondence with the spectrum of the continuous operator. Note that, by ordering the eigenvalues as explained above, the multiplicity of the comparison between eigenvalues of the operator and of the matrix is satisfied at least up to spectral index $l = k a$ (we assume $k a$ is integer). Indeed, given a spectral index $l \le k a$, we identify $(2l+1)$ eigenvalues of $\bar{\mat S}^k$ to be compared with $\lambda^{\op S^k}_l$, each of them eigenvalue of a different block $\bar{\mat S}^k_{p}$ characterized by $|p|\le l$.

The numerical results proposed in the following will focus on polynomial basis functions. In particular, we will consider patch $\uppi$ and pyramid $\uplambda$ basis functions, given by the product of patch and pyramid basis functions of argument $\phi$ and $z$, denoted as $\uppi^\phi$, $\uplambda^\phi$, $\uppi^z$, $\uplambda^z$. 
Further analyses are required to verify the convergence of such bases.
The Fourier coefficients associated to $\uppi_0^\phi$ and $\uplambda_0^\phi$, with expressions \cite{warnick2008numerical}
\begin{equation}
    T_p^\uppi=\left[\frac{\sin(\pi p/M)}{(\pi  p/M)}\right]\quad\text{and}\quad  T_p^\uplambda=\left[\frac{\sin(\pi p/M)}{(\pi p/M)}\right]^2\,,
\end{equation}
equal one for $p = 0$ and equal zero for $p = sM$. This allows for a simplification of the expression of the block $\bar{\mat S}^k_{p=0}$, reading
\begin{equation}
    \bar{\mat S}^k_{p=0} = -\frac{V}{2 a^2}  \sum_{l=0}^\infty \lambda^{\op S^k}_l \mat P_l^{0}\,.
    \label{eqn:Sp0}
\end{equation}

Following similar steps, the eigenvalues of the matrix $\mat N^k$ discretizing the hypersingular operator can be retrieved as the union of the eigenvalues of the $M$ blocks $\bar{\mat N}^k_p$ defined as
\begin{equation}
    \bar{\mat N}^k_p = \frac{V}{2 a^2} \sum_{s=-\infty}^\infty  \sum_{l=|p+sM|}^\infty \lambda^{\op N^k}_l \mat P_l^{(p+sM)} T_{-(p+sM)}F_{(p+sM)}\,.
\end{equation}
Finally, by virtue of the spherical harmonics decomposition of the Dirac delta distribution \cite{abramowitz1964handbook}, the eigenvalues of the Gram matrix $\mat G$ discretizing the identity operator, with eigenvalues $\lambda^{\op I}_l = 1$, are given by the union of the eigenvalues of the blocks
\begin{equation}
    \bar{\mat G}_{p} 
    = \frac{V}{2 a^2} \sum_{s=-\infty}^\infty  \sum_{l=|p+sM|}^\infty \lambda^{\op I}_l \mat P_l^{(p+sM)} T_{-(p+sM)}F_{(p+sM)}\,.
\end{equation}


\section{High-Frequency Analysis of the Spectral Error}
\label{sec:analysis}
As shown in the previous section, the eigenvalues of the BEM matrices discretizing the integral operators $\op S^k$, $\op N^k$ are the union of the eigenvalues of the blocks $\bar{\mat S}^k_p$ and $\bar{\mat N}^k_p$. In particular, up to the spectral index $l = ka$, a clear correspondence with the eigenvalues of the continuous operators $\op S^k$ and $\op N^k$ can be identified, allowing for the definition of a relative spectral error
\begin{equation}
    E_{l,p}^{\op S^k} \coloneq \frac{\hat{\lambda}_{l,p}^{\mat S^k}-{\lambda}_{l}^{\op S^k}}{{\lambda}_{l}^{\op S^k}}\,,\quad
    E_{l,p}^{\op N^k} \coloneq \frac{\hat{\lambda}_{l,p}^{\mat N^k}-{\lambda}_{l}^{\op N^k}}{{\lambda}_{l}^{\op N^k}}\,,
\end{equation}
where we have denoted as $\hat{\lambda}_{l,p}^{\mat S^k}$ and $\hat{\lambda}_{l,p}^{\mat N^k}$ the eigenvalues of the blocks $\bar{\mat S}^k_p$, $\bar{\mat N}^k_p$ associated to the spectral index $l$.

We did not observe, nor did we theoretically find, any orthogonality between $\{\vec t_l^p\}_{l=|p|}^\infty$ and $\{\vec f_l^p\}_{l=|p|}^\infty$ associated to the basis functions considered. Hence, we may assume that the aliasing component of the spectral error \cite{warnick2008numerical}, which here means the component of the spectral error resulting from interferences with high-frequency spectral components of the spectrum of the operator outside the visible range, is given in this case both from the terms corresponding to $s \ne 0$ and from the terms corresponding to $l > |p|+V$. 

It follows that, as in the two-dimensional circular case \cite{giunzioni2024highfrequency}, the high-frequency behavior of the spectral error will be determined by the high-frequency behavior of the spectra of the continuous operators, and, in particular, by the relative high-frequency scaling of some spectral components with respect to others. Note that \cite{abramowitz1964handbook}
\begin{equation}
    j_l(ka) = \sqrt{\frac{\pi}{2ka}} J_{l+\frac{1}{2}}(ka)\,\,,\quad
    h_l^{(2)}(ka) = \sqrt{\frac{\pi}{2ka}} H^{(2)}_{l+\frac{1}{2}}(ka)\,,
\end{equation}
where $J_l$ and $H^{(2)}_l$ are the Bessel and Hankel functions of first and second kind \cite{abramowitz1964handbook}.
By studying the asymptotic expansions of Bessel and Hankel functions, of large argument type (\cite[Section 9.2]{abramowitz1964handbook}) for indices $l \ll ka$ and of large order type (\cite[Section 9.3]{abramowitz1964handbook}) for $l \simeq ka$ and $l \gg ka$, we obtain that $|\lambda_l^{\op S^k}|$ and $|\lambda_l^{\op N^k}|$ in the transition region respectively increase as $(ka)^{1/3}$ and decrease as $(ka)^{-1/3}$ in the high-frequency regime, while they maintain a constant order of magnitude in the hyperbolic and elliptic regions of the spectra. As in the two-dimensional case \cite{giunzioni2024highfrequency}, this translates into an aliasing error expected to decrease as $(ka)^{-1/3}$ and to increase as $(ka)^{1/3}$ in the transition region for the matrices $\mat S^k$ and $\mat N^k$.

\section{Numerical Results}
\label{sec:results}
\begin{figure}
\centerline{\includegraphics[width=1\columnwidth]{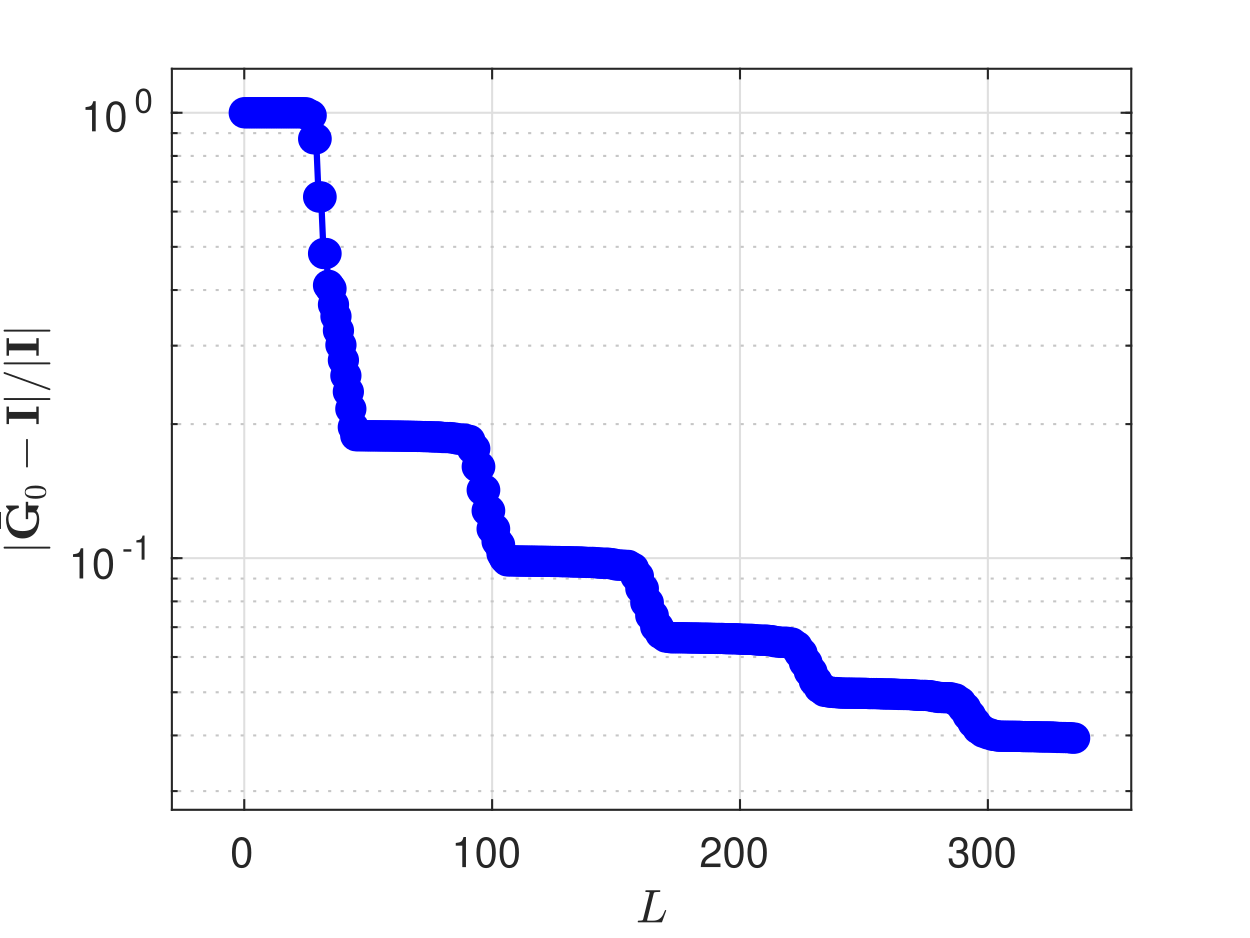}}
\caption{Convergence of the Gram matrix evaluated following our new semi-analytic approach (\cref{eqn:grampatch}) to the expected value as a function of $L$.}
\label{fig:gramConv}
\end{figure}

The numerical results proposed in this section aim at validating the novel semi-analytic approach for the evaluation of the eigenvalues of the BEM matrices. To this end, we compared matrix
\begin{equation}
    \tilde{\bar{\mat G}}_{p=0} = \frac{V}{2 a^2} \sum_{l=0}^L \lambda^{\op I}_l \mat P_l^{0}\,,
    \label{eqn:grampatch}
\end{equation}
resulting from the discretization by means of patch functions, with the identity matrix, which is the expected result for the Gram matrix for $L\rightarrow\infty$, given the uniform parametrization of the sphere. The convergence of $|\tilde{\bar{\mat G}}_{p=0}-\mat I|/|\mat I|$ to zero as a function of $L$ is shown in Figure \ref{fig:gramConv}, validating our approach.

\begin{figure}
\subfloat[\label{subfig-1:spectraS0_k30}]{%
\includegraphics[width=1\columnwidth]{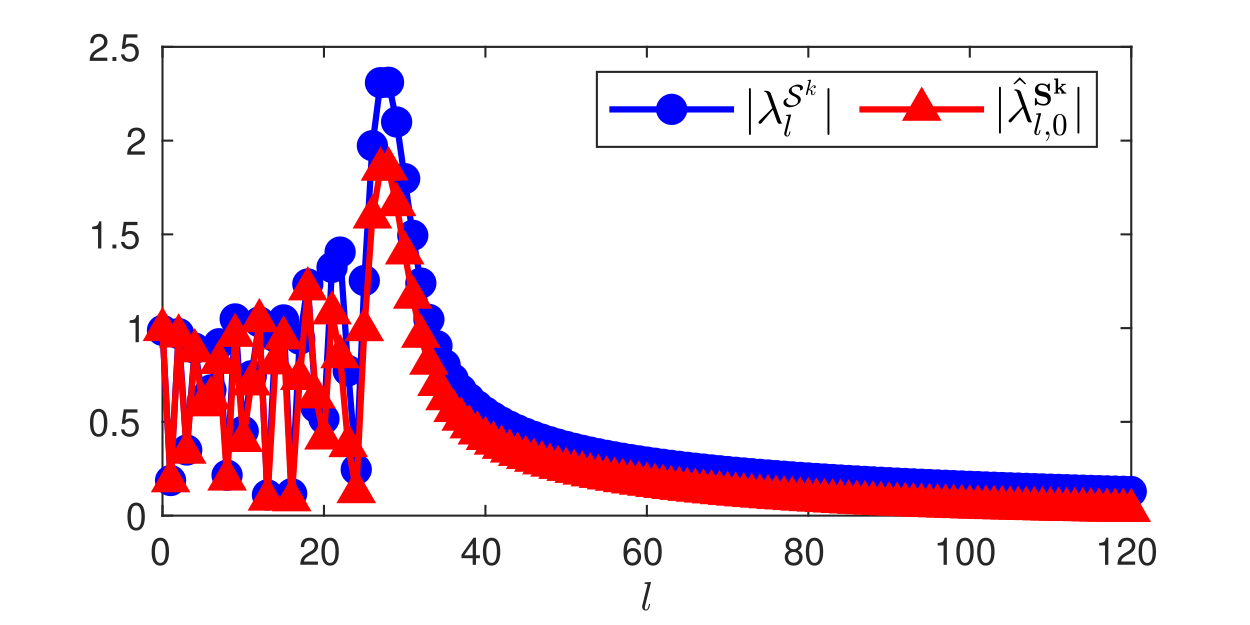}
}
\hfill
\subfloat[\label{subfig-2:spectraS0_k30}]{%
   \includegraphics[width=1\columnwidth]{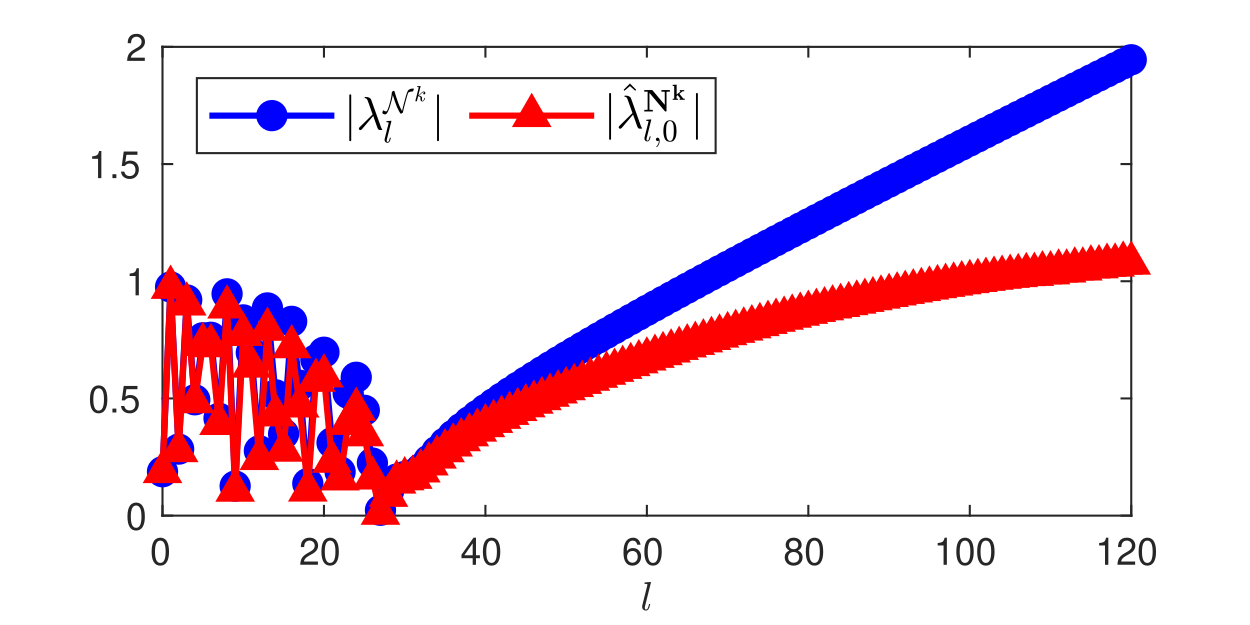}
}
\caption{Eigenvalues of the BEM matrices of the single-layer (a) and hypersingular (b) operators discretized with pyramid basis functions evaluated following the semi-analytic approach proposed in \Cref{sec:eigenvalues}: comparison with the spectra of the operators. }
\label{fig:spectra}
\end{figure}

We then employed the proposed strategy to evaluate the eigenvalues of the single-layer and hypersingular operator matrices (eigenvalues of $\bar{\mat S}^k_{p=0}$ \eqref{eqn:Sp0} and $\bar{\mat N}^k_{p=0}$) discretized by means of pyramid functions at the frequency $ka = 30$, shown in Figure \ref{fig:spectra}.

\section{Conclusions}
\label{sec:conclusion}
We proposed a new semi-analytic approach for the spectral characterization of the BEM matrices discretizing integral operators over the sphere. The high-frequency analysis of the spectral relative difference between matrices and operators reveals an increasing behavior as $(ka)^{1/3}$ of the spectral error of one of the operators considered, suggesting the possibility of an increasing solution error in the high-frequency regime. 

\section*{Acknowledgements}
This work was supported by the European Innovation Council (EIC) through the European Union’s Horizon Europe research Programme under Grant 101046748 (Project CEREBRO) and by the European Union – Next Generation EU within the PNRR project ``Multiscale modeling and Engineering Applications'' of the Italian National Center for HPC, Big Data and Quantum Computing (Spoke 6) – PNRR M4C2, Investimento 1.4 - Avviso n. 3138 del 16/12/2021 - CN00000013 National Centre for HPC, Big Data and Quantum Computing (HPC) - CUP E13C22000990001.

{ \printbibliography}

\end{document}